# Time, Quantum Mechanics, and Probability

Simon Saunders

ABSTRACT. A variety of ideas arisiüg in decoherence theory, and in the ongoing debate over Everett's relative-state theory, can be linked to issues in relativity theory and the philosophy of time, specifically the relational theory of tense and of identity over time. These have been systematically presented in companion papers (Saunders 1995, 1996a); in what follows we shall consider the same circle of ideas, but specifically in relation to the interpretation of probability, and its identification with relations in the Hilbert space norm. The familiar objection that Everett's approach yields probabilities different from quantum mechanics is easily dealt with. The more fundamental question is how to interpret these probabilities consistent with the relational theory of change, and the relational theory of identity over time. I shall show that the relational theory needs nothing more than the physical, minimal criterion of identity as defined by Everett's theory, and that this can be transparently interpreted in terms of the ordinary notion of the chance occurrence of an event, as witnessed in the present. It is in this sense that the theory has empirical content.

## 1. Introduction

Following Everett, there are reasons to suppose that quantum mechanics can be understood as a universal physical theory, which operates with only the unitary dynamics. Since the unitary theory is not supplemented with any principle whereby one component of state is singled out at the expense of all others, it is committed to a variety of modal realism: in some sense all physical possibilities are realized. For many this is good enough reason to reject Everett's ideas. But we are between a rock and a hard place; the alternatives, state-reduction and hidden-variable theories, require revision of physical principles even outside of quantum mechanics, in particular the principle of relativity. If unitary methods can really deliver, the more conservative course is to make do with them, and with the physics as is.

But do they deliver? Two problems remains outstanding: the *preferred basis problem*, the problem of accounting for the partitioning up of possibilities, in accordance with the various possible positions of massive systems; and the *problem of probability*. Progress has been made with the preferred-basis problem. In part this is due to a better understanding of decoherence theory (the "effective" washing out of interference effects), and in part it is due to a shift in the nature of the problem. The latter is from a philosophical point of view the most interesting: the question is no longer: What is the space of possibilities? - but: What is the space of possibilities *in which we are located*? That we are ourselves constituted of processes of a very special sort is in part explanatory;[1] because our nature is in part contingent, a matter of evolutionary circumstance, the demand for explanation is blunted. For a systematic treatment, I refer to a companion paper (Saunders 1995).

If so the problem of probability is all the more urgent[2]. This is our business in what follows. The problem amounts to this: it seems that the concept of probability can only apply to a situation given that only one, say, *x*, out of a range of alternative possibilities *y*,*z*,..., is true, or is realized, or actually occurs, so as to exclude all the others; precisely what Everett denies.

This reasoning is less clear than it seems, but we can break it down into steps. It is supposed that

(A1) *x* has probability *p*.

Typically, by this we mean:

(A2) *x* will happen with probability *p* .

If probability is to have any application, there had better be a range of alternative possibilities $x, y, \ldots, z$. If *x* does happen with probability *p*, then the alternatives are surely excluded:

(A3) if *x* happens *y* does not happen .

(A3) can also be given an explicit temporal sense, as in the transition from (A1) to (A2).

All these are obvious platitudes. But (A3) is denied by Everett, who says that in some sense

(A4) *x* happens and *y* happens.

Without (A3) we cannot make sense of statements like (A2) and (A1). So there can be no probability. Everett's approach was anyway incredible, committed as it is to an extravagant ontology; now we see it is internally inconsistent as well.

On the contrary, I claim that the problem of probability can be fully resolved in Everett's framework. What is needed is a thorough-going relativisation of physical modal attributes, specifically of value-definiteness and probability, viewed as an extension of the relativization of tense familiar to classical physics. I have elsewhere argued that classical relativity theory, both special and, barring highly idealized cases, general relativity, force a relational theory of tense (Saunders 1996a); in consequence this much must be taken over to relativistic quantum theory. But then there are few new problems, that arise if we extend the relational approach to tense to other modal attributes, in particular to possibility and probability (and thereby to necessity and determinism as limiting cases), or so I claim.

The outstanding question concerns the choice of basis, or equivalently, the choice of decoherent history space. I shall assume throughout that this is fixed, at least in the approximate sense consistent with the condition of medium decoherence (in the sense of Gell-Mann and Hartle). Here we are concerned with simple and intuitive difficulties concerning the interpretation of probability.

## 2. Relational Time

The relational approach is from the outset concerned with extending the relational account of tense to modal attributes.[3] This is a radical move, so it is easy to lose sight of what is already involved in the more familiar step, the relativization of tense, particularly as it figures in relativity theory. This early and familiar step is in some ways the more significant.

First and foremost time is geometrized; we work exclusively with a 4-dimensional manifold, and with its various metrical structures. As a result we do violence to a number of intuitive notions bound up with time, specifically tense, the “flow” of time, and identity over time. For example, the geometrical account of identity over time - in terms of the spatiotemporal continuity and structural similarities among events - does not seem to express the concept of a thing which can exist at different times. The problem is most obvious at the level of personal identity: I suppose that however I may change from one day to the next, there is an irreducible sense in which I am one and the same: that the whole of me is at each time, and yet is the very same at all these times.

This doctrine, however murky and apparently contradictory, undeniably plays a role in our ordinary lives. Most of the debates in the philosophical literature focus upon this, and with it the irreducibly tensed notion of reality as *what exists* (as what exists *now*, what is therefore a 3- dimensional world at one time). If, conversely, we do accept the adequacy of the 4-dimensional representation of the world - as I say we should - then it may be that certain of our intuitive notions of change and identity have to be jettisoned, no matter that they are commonplace.

Consider now our understanding of space. We suppose that we have an immediate grasp of a 3-dimensional world, of 3-dimensional space as a whole, by simple extrapolation from the perception of things in space about us. The idea of 3-dimensional physical space is supposed to be more or less self-evident. But that is certainly false, for if it is physical space then it had better be a 3-dimensional space at one time, i.e. a space-like hypersurface. Specifying this requires a choice of simultaneity relation, a moderately complicated procedure. This is not something of which we have any intuitive idea or perceptual datum. We ordinarily think that what we see at a glance co-exists, but from a naturalized view, we see events distributed over a 3-dimensional hyper-surface in 4-dimensional space-time, our past light cone, a space which is isotropic but neither space-like nor homogeneous. Is this physical space? If so, it is as anthropocentric a concept of space as any due to Aristotle or Ptolemy.

The two examples are related, for the simple reason that “the now”, to play the part of a fundamental metaphysical category, along with “the real”, and “the actual”, had better, at each instant, be something unique and inter-subjective. Indeed, it had better be something of universal significance. But it is the express claim of our best theory of space and time - the theory of relativity - that there is *no* such unique and universal object, “the now”, or series of such, (for that would imply the existence of a uniquely distinguished space-time foliation).[4] So much the worse, then, for the metaphysical categories based upon it. The most important of these is “substance”, the substratum of changing attributes which does not itself change. Without this the metaphysical notion of identity over time, as something different from “genidentity” or similar notions, derived from criteria of physical spatio-temporal continuity, goes by the board; for it is by virtue of being one and the same thing, at different times, that a thing can possess different attributes or determinations, at those times, and yet *be the same thing*. It is through being one and the same “mind” - substance *par excellence* - that we count ourselves identically the same at different times.

What does all this have to do with the problem of probability? Simply this: the concepts of identity and substance play a tacit role in setting up the problem. It is not the concept of probability *per se* which is at issue.

# 3. Relational Probability

On occasion the concepts of identity and substance play a more explicit role. For example, according to Albert and Loewer

> .... the cost of surrendering the “trans-temporal identity of minds” would seem to be that we can no longer make sense of statements like “the probability that I will observe spin up on measurement is *p*” since such statements seem to presuppose that it makes sense to talk of a single mind persisting through time. (Albert and Loewer, 1988, p.211).

But granted a basic premise of the relational theory - the adequacy of the 4-dimensional framework - this cuts no ice at all. There are too many examples of conflicts of this kind, in ordinary language expressions, already at the level of tense.

Given the relativism of tense, the everyday use of terms such as “happens”, ”will happen”, etc., is that there is always a tacit context, the context of use, and that this context must be made explicit. The same applies to probability. Given the context of (A1), denote *z*, we have:

(R1) *x* has probability *p* relative to *z* .

(A2) adds to this the fact that x is temporally related to z (and is in fact later):

(R2) $x$ is later than $z$ and has probability $p$ relative to $z$ .

In the case of (A3), we may allow that $x$ and $y$ are relativized to a single third event $z$:

(R3) if $x$ happens relative to $z$ then $y$ doesn't happen relative to $z$

but it would beg the question to suppose that $z = z'$ , on rewriting (A4) as:

(R4) $x$ happens relative to $z$ and $y$ happens relative to $z'$

since Everett expressly required that each of two distinct and incompatible experimental outcomes be correlated with *distinct and incompatible* states of the observer, or recording apparatus, i.e. that $z$ and $z'$ are not to be identified.[5]

So much for events referred to in impersonal terms. Considering again Albert and Loewer's remark, we see they shift to a first-person version of (A2):

(A2′) the probability that I'll observe spin up on measurement is $p$ .

(A2′) has a more complicated relational structure. We build up to a relational reading of it by stages. First, given

the probability that I'll observe spin up on measurement at $t_2$ is $p$

who does the observing, I at $t_2$ or I at some earlier time? At $t_2$ of course:

the probability that I at $t_2$ observes spin up on measurement at $t_2$ is $p$ .

But what is the tacit context of "probability that" - is this the probability of observing spin up relative to the state of affairs at $t_2$? Surely not; given the state of affairs at $t_2$, the matter is settled. Rather:

(R2′) the probability relative to I at $t_1$ that I at $t_2$ observes spin up on measurement at $t_2$ is $p$.

We can make the same sort of sense of (A2′) as (A2); the difference is that the notion of personal identity is more fully in view. How is "I at $t_1$" related to "I at $t_2$"? The answer is that they are related by the Hilbert-space norm, and that this is probability; that likewise $x$ is related to $y$ by a temporal interval, and that this is time.

But at each time $t_1$ there are many possibilities for "I at $t_2$ ", to many of which I at $t_1$ am related. For this reason Albert and Loewer insist that the relation be 1:1; their concern is with sentences of the form

(A2*) I will observe spin up .

As it stands this does more than assert that I at $t_1$ am earlier than I at $t_2$ , who observes spin up; there is the further implication that there is no doubt about the matter, that I at $t_1$ will certainty be I at $t_2$ (that I at $t_1$ am identical with I at $t_2$ ). We can indeed cash this out in terms of a deterministic dynamics, and by insisting that there is a unique world-line stretching from I at $t_1$ to I at $t_2$ ; but now it is not the concept of probability that has led to a unique criterion of identity over time, but rather certainty or determinism. Albert and Loewer can hardly insist that in order for the concept of probability to make sense, there must exist a unique and determinate future, when so many have argued that non-epistemic probability only makes sense given that the future is *not* something determinate (see, e.g. Maxwell 1985).

Many philosophers take the peculiarities of the various relational readings of these sentences as evidence for the failings of relationalism; but equally, we could conclude that our ordinary conception of change is muddled, and involves much else besides physics. How are we to picture the process of probabilistic becoming? I say that it is to be understood as a system of relations, the same here as with deterministic becoming, in which notions of space-time and probability function as primitives. The "problem of probability", so-called, is the problem of how to provide something more. But we have learned to live with this lacuna, in the deterministic case, and we can do the same in quantum mechanics.

Evidently we need to consider more detailed objections, and at a leisurely pace. We begin with a series of simple models. Only later shall we consider more realistic theories.

# 4. Examples

Consider experiments performed at times $t_1, \ldots. t_N$, each with $M$ possible outcomes. We can represent all these outcomes by means of a tree-diagram with a preferred orientation, so that each vertex has one incoming and $M$ outgoing lines[6]. Each line, connecting two vertices, we suppose is labelled with the transition probabilities between the state preparation and measurement outcome represented by the two vertices. A *history* is a continuous sequence of such lines, with vertices totally ordered by the time. Everett supposed that probability theory enters into quantum mechanics as a measure over the space of histories, in parallel to classical statistical mechanics, where probability is defined in terms of a measure on phase space. In this (Everett 1957, 1973) he was brief but explicit. Insofar as he acknowledged that something more was required, it was that the particular choice of measure should be justified, and shown to be consistent with Born's rule. This is what he tried to do. No more than this, he remarked, can be demanded of any probability theory, neither quantum mechanics nor classical physics.

In one sense this is quite right, as I shall argue later. But it is not as it stands a response to the problem of probability. Indeed, the limitations of purely formal arguments became more apparent once Everett's program is carried through. The problem of how to justify the choice of measure was independently raised and solved by Gleason's celebrated result that same year (Gleason 1957): the only normalizable, countably additive measures over disjoint sets of projections are those given by density matrices, or in Everett's case, by the universal state[7]. The connection with statistics was also clarified; the quantum mechanical Bernouilli theorem, as first formulated by Finkelstein in 1963, states that those histories, which encode anomalous relative frequencies on $N$ repetitions of a given experiment, have vanishing measure in the limit as $N$ becomes infinite[8].

The question that remains is how to justify the interpretation of this measure as probability, when the limit is not reached (and in real life it *never is* reached). We cannot reduce probability to the $\{1,0\}$ case, in the context of physically realistic statistics, involving finite populations. As a first step, then, we will consider strictly finite models, for $N$ repetitions of experiments with $M$ possible outcomes, with $M$ and $N$ finite.

## 4.1 Many Worlds

The problem which then arises, in the many-worlds approach of DeWitt and Deutsch, is that:

> ....it is extremely difficult to see what significance [Everett's] measure can have when its implications are completely contradicted by a simple count of the worlds involved, worlds that Everett's own work assures us must all be on the same footing. (Graham 1973 p.236).

Graham, a student of DeWitt, has just computed the proportion of histories of $N$ yes-no outcomes (vertices with $M = 2$), each with probability $p$ for a positive result, with $K$ positive outcomes, i.e. with relative frequency $K/N$. The number is $N!/(N-K)!K!$, independent of $p$; it is a maximum for $K/N \approx 1/2$, so the greater number of histories encode statistics that are only correct if $p = 1/2$. There is no connection between the numbers of histories with the "right" statistics, and the probability $p$ for each trial, which is supposed to be approximately equal to the relative frequency. Hence Graham's complaint. It has been echoed by many others since:

> The many-worlds theory is incoherent for reasons which have been often pointed out: since there are no frequencies in the theory there is nothing for the numerical predictions of quantum theory to mean. This fact is often disguised by the choice of fortuitous examples. A typical Schrödinger-cat apparatus is designed to yield a 50% probability for each of two results, so the 'splitting' of the universe in two seems to correspond to the probabilities. But the device could equally be designed to yield a 99% probability of one result and 1% probability for the other. Again the world 'splits' in two; wherein lies the difference between this case and the last? (Maudlin, 1994, p.5).

Maudlin says enough here to see the outlines of an argument. It is claimed that ratios in the numbers of histories of certain sorts defines the probability of that sort of history. Why make this supposition? Evidently it would follow if all histories were equiprobable; is there reason to think that they are?

What reason there is appears to derive from the picture of "splitting": were the splitting perfectly symmetric, that would be grounds to suppose each of the two branches equiprobable. But then, what would count as a non-symmetric splitting? A simple retort is it is the value of $p$ that determines whether or not the splitting is symmetric; by all means make these values explicit in a tree-diagram, by the thickness of the lines. But against this Maudlin is saying that something more is needed, than an artifact of some graphical representation; what is needed is an understanding of *how* these numbers are to be understood as probabilities.

In response to this challenge some have extended the metaphor; thus Lockwood, with talk of an internal dimension "orthogonal" to space and to time (Lockwood 1989, 1996). But following Deutsch, he and others suppose that the notion of measure, to have any meaning defined on this internal space, must consist in a measure of a new plurality of worlds (or, in Lockwood's case, of "perspectives"); that we must introduce a further collective, all of whose members are exactly the same, prior to the splitting. Only then can we interpret the probabilities of each branch, following the split, in terms of the proportion of worlds that end up in that branch.

This is an *ad hoc* device, providing, as it stands, no more than an aid to the imagination.[9] We shall make no use of it here. No more is it needed; for consider the analogous situation in state-reduction theories; (there we find the "problem of tails"; at each instant, even allowing for a mechanism of state reduction, we still have a superposition of states, associated, by the conventional eigenvalue-eigenvector link, with distinct pointer positions at $x$ and $y$, with states $|x>$ and $|y>$. The only difference, from the usual unitary dynamics, is that the *amplitude* of the one is made enormously greater than the other (which one depending on an irreducibly stochastic element to the dynamics). But why should a mere difference in *numbers* make for a physical difference, between outcome $x$ and $y$?

The problem is not avoided by abandoning the eigenvalue-eigenvector link, and supposing that such a superposition exactly represents the pointer as localized at $x$; for why does *that* superposition, of the form:

$$\Psi = c_x \mid x > + c_y \mid y >, \quad \mid c_x \mid^2 \gg \mid c_y \mid^2 \tag{1}$$

describe the pointer at $x$ rather than at $y$, whereas that with $\mid c_x \mid^2 \ll \mid c_y \mid^2$ describes the pointer at $y$ rather than $x$? The answer, clearly, is that *by fiat*, the norms tell us whether the object is localized in one place rather than another. But in that case why can't we say the same in the Everett theory, that *by fiat*, the transition probabilities tell us which component of state is probable rather than improbable? It is hard to see why the one is acceptable, but not the other. The norms that figure in state reduction theories are not, after all, a matter of "size", or of "density", of something familiar to classical physics. Nor do they mark a "degree of reality", or the partly real; the event $x$ is wholly real, insofar as that is the position of the pointer at the end of an experiment, in accordance with Eq.(1), for all that the only difference between the state describing the pointer at $x$, and that describing it at $y$, lies in the relative weights $|| c_x |^2, | c_y |^2$, and that neither is ever zero. So it is not that we have any *antecedent* understanding of the meaning of these norms, whereas, in the case of the transition probabilities, there is none; in both cases these quantities are *sui generis*; neither can be reduced to something else, or explained in terms of something antecedently understood.

But there does remain a distinction between the two cases. Differences in the Hilbert space norms, in Eq.(1), makes for different states of affairs, but at least, in the case of state-reduction theory, one or the other norm dominates when it comes to macroscopic states of affairs. At each time, we have a clear conception of what these states describe - the positions of pointers. What happens when the norms are comparable in magnitude? What sort of world, or worlds, do we have in mind?

But we have not as yet got clear about tense; what has probability $p$ lies in general in the future. Indeed, it is the relational approach which fits better with intuitive ideas in this regard: events in the future, having relationship $p$ with the present, close to neither 1 nor 0, are *indeterminate*; and how else should we think of the indeterminacy or "openness" of the future? That future possibilities are not yet settled? The transition probabilities, relations in the Hilbert space norm between future events and the present, express the degrees of this indeterminacy[10].

Indeterminacy lies in the future. In physics probability habitually involves time. Typically, the concept of probability applies to states of affairs qua future, in relation to the present. Correspondingly, probabilities are conditional, they are *de facto* relations. The point is even clearer if we formulate quantum mechanics in terms of path integrals: what are calculated are transition amplitudes.

What do these transition probabilities govern? Obviously, probabilistic happenings and comings-about. Likewise temporal relations in the deterministic case govern change and becoming. There are objections in both cases to their expression in geometrical and relational terms, but at bottom they are the same. How are we to understand probabilistic becoming, as something other than a system of relations? The analogous problem, in the deterministic case, is the one that is familiar to philosophy. My answer is to simply rest with this connection: the two problems are the same. We cannot understand deterministic becoming, in other than relational terms, and neither can we understand indeterministic becoming, in other than relational terms. We are no worse off in quantum mechanics than in classical relativistic physics. This cannot be a decisive objection to Everett's approach.

## 4.2 Many Minds

It may be objected that Maudlin grants too much; that even in the case of symmetric division, we cannot meaningfully suppose that the probability of each branch is one half. Arguments to this effect have been formulated by Loewer, in defense of the mechanisms postulated by the "many-minds" theory of Albert and Loewer (1988), and in explicit opposition to what Loewer calls the "IMV" or "instantaneous minds view". This is the view that there is no fact of the matter, as to what minds at one time will become at later times.

Talk of purely mentalistic items, “minds”, in place of their material counterparts, “worlds”, makes for better focus on the relation of physical probability to degrees of belief, and to first-person expectations, but otherwise has little to recommend it. On the other hand, Loewer’s criticisms of the IMV apply equally to relationalism:

> Prior to measuring the *x*-spin of a *z*-spin electron, a rational observer who believes the IMV ought not to have a degree of belief of 1/2 that she will observe spin up. Either she will think that this degree of belief is 0 because she will not exist at the later time or, if she identifies herself with all the minds associated with her brain at the end of the measurement, she will believe that at the conclusion of the experiment she will certainly perceive that *x*-spin is up and also she will believe that *x*-spin is down and so assign a degree of belief of 1 to each of these. (Loewer 1996 p.230).

Loewer supposes that the belief is about what the observer will see after the measurement. “The observer” at what time, and on what outcome? Evidently the participle ‘she’ is allowed to shift indiscriminately from one context to another: from she$_1$ prior to the experiment, at time $t_1$; to she$_2^{\uparrow}$ at $t_2$ following the experiment, observing outcome spin-up, and to she$_2^{\downarrow}$ at $t_2$ , observing outcome spin-down. The question, surely, is what will she see? (There is no doubt about she$_2^{\uparrow}$ and she$_2^{\downarrow}$.) Specifically, her degree of belief, at $t_1$, concerns whether she$_1$ will become she$_2^{\uparrow}$ or she$_2^{\downarrow}$; it concerns her anticipation of what is to come.

Loewer does not appear to recognize that this is what her$_1$ belief concerns; or if he does, he seems to suppose that she$_1$ should anticipate nothing at all, because she$_1$ does not exist at $t_2$. As an alternative to this implausible suggestion, he allows only that she$_1$ may consider, not who she$_1$ will be, but what exists at $t_2$ - about which she$_1$ has no doubt whatsoever.

This is no alternative; it is a different matter entirely. The genuine alternatives appear to be these: either she$_1$ anticipates being both she$_2^{\uparrow}$ and she$_2^{\downarrow}$, some kind of composite; or else she$_1$ anticipates being either she$_2^{\uparrow}$ or, in the exclusive sense, she$_2^{\downarrow}$.

Nothing, both, or else just one of them? I have said that the first option is implausible: certainly it is not because she$_1$ does not exist at time $t_2$, that she should expect nothing at all; the antecedent is true for ordinary deterministic change, in classical physics. As for the second option, it is straightforwardly inconsistent with the evidence; she$_2^{\uparrow}$ and she$_2^{\downarrow}$ do not speak in unison; they do not share a single mind; they witness different events. We do not know what it is to anticipate observing incompatible outcomes, at a single time. There remains only the third alternative: she anticipates being one of she$_2^{\uparrow}$ and she$_2^{\downarrow}$, but not both at once.

We come to the crux of the matter. Is this proposal intelligible? Am I entitled to the belief, at $t_1$, that I will observe spin-up or spin-down, but not both?

I say again that any argument to the contrary had better stop short of applying to the deterministic case as well; it had better not imply that I at *t* cannot expect to be anything other than what I at *t* already am. The burden of argument is, moreover, with Loewer; it is he who proposes to supplement the physics, introducing new and fundamental equations of motion (the stochastic dynamics, governing the dynamics at the level of “minds”). It is the relational approach which is conservative, and makes do with physics as is.

Neither is it routinely required, of a physical theory, that a proof be given that we are entitled to interpret it in a particular way; it is anyway unclear as to what could count as such a proof. Normally it is enough that the theory can be subjected to empirical test and confirmation; quantum mechanic can certainly be *applied*, on the understanding that relations in the Hilbert-space norm count as probability (with use of the projection postulate where appropriate, i.e. relativizing to the context of measurement). It is not as though the experimenter will need to understand something more, a philosophical *argument*, for example, before doing an experiment.

To that it can be countered: the problem of measurement has always been a *philosophical* problem; experimentalists never have needed to resolve it (or understand it). But there I disagree. At the present point we are talking of the applicability of the formalism; an important aspect of the problem of measurement is not philosophical, except in the broadest sense of the term, for we want to apply quantum mechanics to the early universe, and to cosmology (where there is no *external* observer). Granted that the relational approach offers a framework for quantum cosmology, that is proof enough that we have made a decisive step forward in solving the problem of measurement, even if there *do* remain aspects of the physical interpretation, and of the formal representation of probability, that are at odds with standard intuitions. It may be that our usual intuitions stand in need of correction.

Leower’s task is harder than he thinks. For he has to demonstrate the internal inconsistency of the relational approach; it is not enough for him to show that intuitive views on the nature of change and becoming are violated, for that is already true of classical relativity theory. And there is nothing inconsistent in the view that I expect to become either $\text{he}_2^{\uparrow}$ or $\text{he}_2^{\downarrow}$ (but not both), in the sense that I at $t_1$ look forward to $\text{his}_2^{\uparrow}$ memories and outlook, or to $\text{his}_2^{\downarrow}$ , but not to having both at once; whilst acknowledging that both will count me at $t_1$ as their common earlier self.

On any naturalist view of human behavior, the view is entirely reasonable. For example, at the level of *linguistic behavior*, $\text{he}_2^{\uparrow}$ and $\text{he}_2^{\downarrow}$ will both declare that $\text{his}_1$ expectation of becoming the one or the other, but not both, was fully satisfied. With transition probability close to 1, moreover, they will declare that the observed statistics (of, say, an experiment running in tandem, involving many particles), is in accordance with the relations *p*; in other words, with probability close to *p*, the relative frequencies approximately equal what quantum mechanics says they will be. Likewise when viewed in evolutionary terms; those successors, whose expectations are not tailored to the quantum mechanical quantities, the relational probabilities, will have probability close to zero of surviving.

Loewer’s argument in fact depends on tacit and tendentious intuitions, precisely those already called into question in the relational account of tense. In effect he insists that if $\text{she}_1$ becomes $\text{she}_2^{\uparrow}$, then $\text{she}_1$ does not become $\text{she}_2^{\downarrow}$; the latter is in consequence a mere automaton, no matter that $\text{she}_2^{\downarrow}$ also declares that $\text{she}_2^{\downarrow}$ was $\text{she}_1$; and this is surely unacceptable. But if, per impossibile, $\text{she}_1$ does become she$_2^{\downarrow}$, despite the fact that she$_1$ becomes she$_2^{\uparrow}$, then surely she will have become both, and it follows that she has every right to expect to become both with certainty; and again this is unacceptable. Since the two cases are exhaustive, the IMV is unacceptable, period.

Underlying this reasoning is the old picture of persistence. Rather, “to become”, as in “she$_1$ becomes she$_2^{\uparrow}$”, is for she$_1$ to stand in a certain relation to she$_2^{\uparrow}$, and there is nothing in this relation that requires it to be 1:1. Of course, if a thing, and the whole of that thing, passes from one moment to the next, where it is wholly contained, it is hard to understand how it might also be at a *different* moment, or witnessing a *different* outcome of measurement; but this is a philosophical picture we have already called into question in the deterministic and classical case.

But the picture of passage through time is tenacious. Witness Weyl’s remark, in an attempt to dislodge it:

> The world is, it does not happen. Only to the gaze of my consciousness, crawling up the life-line of my body, does the world fleetingly come to life. (Weyl 1949 p.116.)

With that we can run a version of Loewer’s argument. If my consciousness crawls up the life-line of my body, then it departs from one time, $t_1$ and arrives at another $t_2$ ; in which case my body at $t_1$ has no consciousness, it is a kind of automaton, and not every time is treated on a par. But if it does not, then my consciousness is at both times at once, an absurdity. In contrast, on the relationalist account, the movement of consciousness is already described by the life-line, in terms of the relations among its parts; nothing crawls up my life-line, my life-line already depicts change.As for the second horn of the dillema, consciousness is at both times, but not both times at once. To be at both times is being at the first time, *and then*, i.e. to stand in a temporal relation, to being at the second time.

The relational theory is in these respects fully consistent. It was anyway implausible to insist that the notion of probability in this case is incoherent, for it corresponds to a yes-no experiment with p=1/2; this is the one case - equiprobability - which proponents and critics of Everett are agreed *does* make sense, Graham and Maudlin included.

## 4.3 Many Selves

I have argued that objections to the relational account of probability apply equally to the relational account of tense, in the classical case, and amount to an expression of a familiar philosophical position in that context. If this is all that is left to the problem of measurement - philosophical difficulties familiar to the 4-dimensional viewpoint of special and general relativity - then that is progress of sorts; it is no longer the physical theory which is in doubt.

But the point is a stronger one. Not only do related arguments apply to the debate over tense, but essentially the same arguments apply to a familiar puzzle of personal identity.

The puzzle supposes that, by what ever means, persons can be symmetrically split into two, without any psychological disturbance or loss of memory. Let A be subject to this operation, and of the two that survive him, let the first to awake be declared A's legal heir. Should A expect to retain his property, or find himself dispossessed? The scenario is obviously fanciful; nothing like it is likely to be medically possible in the foreseeable future. Does that matter? Parfit, who has made extensive appeal to such thought experiments, was concerned with questions of moral philosophy (Parfit 1984), the focus of most of the subsequent debates on the subject. There the question of feasibility has been a bone of contention. But here we are concerned with any kind of observer; we can consider more simple forms of life, or ones with more pronounced lateral symmetry than our own[11]; or even artificial life.

For convenience we assume perfect symmetry in the process of division itself (during recovery, we can suppose, small asymmetries are introduced, so that one wakes before the other). With that if A anticipates anything at all, his concerns will be equally balanced between $A^+$ and $A^-$ . A surely will suppose that he will survive, in the case where there is only a single successor; why should he expect oblivion, if now there are two? And since $A^+$ and $A^-$ become quite different persons, he can hardly look forward to being some sort of collective mind. Neither is there reason for him to favor the one over the other - not, at least, before the first of them awakes - since by supposition the process of division is symmetric. So he should expect to become one or the other with equal likelihood; he should view them as equiprobable.

The similarities between this and the quantum mechanical case are evident. We see further that Loewer's objections, directed at the IMV, apply to it too. In fact Loewer formulated three criteria for any meaningful application of the concept of probability. I shall state them in full:

> L1: If it is rational for A to assign a probability of *p* to a future event *E* then there must be some matter of fact (e.g. whether or not *E* will occur) of which A is ignorant.
>
> L2: If A believes that the probability of an event *E* on experiment *e* is *p*, then it is rational for A to believe that on many independent repetitions of *e* the frequency of *E* will be approximately *p*.
>
> L3: If the probability of *E* on *e* is *p* then it is possible that on many independent repetitions of e the frequency of *E* will be approximately *p*.

The focus of his concern is clear. If a process of division of worlds, or minds (or in the present context, the biological self) is fully deterministic, it seems that A knows everything there is to know, so L1 is violated. And it is obvious that the total frequency of *E*, over all the branches, will be independent of *p* (this was Graham's reasoning), violating L3. And in case there is not in fact any real branching, or process of division, but only the appearance of it, we may pose the same difficulty at the level of A's beliefs and rational behavior, in which case L2 is violated.

Let us take them one by one, in the context of the dividing self. Is L1 satisfied in this case, with $p = 1/2$? Here is a matter of fact about which A is ignorant: whether he will be dispossessed. Indeed, as a matter of principle, A must be ignorant on this score, for there is no fact of the matter, period. But isn't this to deny what L1 precisely requires, namely that there is such a fact? But a fact for whom, and at what time? Prior to the division, at time $t_1$, he (denote $A_1$ ) does not know what to expect at later times $t_2$ ; there seems that there is no knowledge that could possibly bear on the matter. But after the division, $A_2^+$, on awakening first, will have no such uncertainty; he knows he retains $A_1$'s legal identity, with all his assets and dues. Neither is $A_2^-$ ignorant of anything on this score; for both of his successors a point on which each was ignorant - each recalling very well $A_1$'s uncertainty - has been resolved. One will be contented, the other express regret - it is not quite clear which will be which - but in either case, the matter will have been settled. Each will have learnt something new.

It is no use for Loewer to argue that all this was known in advance to $A_1$; if, at time $t_1$ , his claim is that $A_1$ should not be uncertain as to what was to come, that, indeed, $A_1$ should have no expectations as to his personal fate, that he should expect oblivion. But that is hardly obvious, or even plausible. Neither can he urge that $A_1$'s successors have learnt nothing new; after all, each will *say*, at time $t_2$, that his uncertainty has been resolved. The same objection applies if it is insisted that $A_1$ should anticipate being the both of his successors, of being them at once. For were he to heed this advice, by whatever extravagant act of the imagination, $A_2^+$ will then report that $A_1$ was mistaken; and likewise $A_2^-$ .

No doubt I have played on ambiguities in the way in which L1 is formulated, allowing each of $A_2^+$ , $A_2^-$ to appropriate $A_1$'s "identity" (whatever that is). My point is that the ambiguities are there in our ordinary use of words. Might L1 be reformulated to remove them? Surely it can; for example, require that there is already, at time $t_1$, a matter of fact about some future event *E*, of which A is ignorant. Call this requirement L1*. With that the scenario of personal division cannot count as a case of probability. Alas, neither will Albert and Loewer's stochastic theory count as probability either; nor does L1* have any inherent plausibility. There are other versions of L1 that will rule only against the relational theory, but they amount to little less than the expression of rival theories.

As for Loewer's remaining criteria, L2 and L3 are obviously satisfied if $p = 1/2$. The interesting question is whether they are satisfied for unequal probabilities. To this end consider successive divisions, as illustrated in Fig.1. Grant that the successors on each division are equiprobable; in that case the probability of E relative to A is 1/4. On many repetitions of the 2-stage divisions, does it follow that A should believe that the frequency of events of type E will be 1/4? Is L2 satisfied?

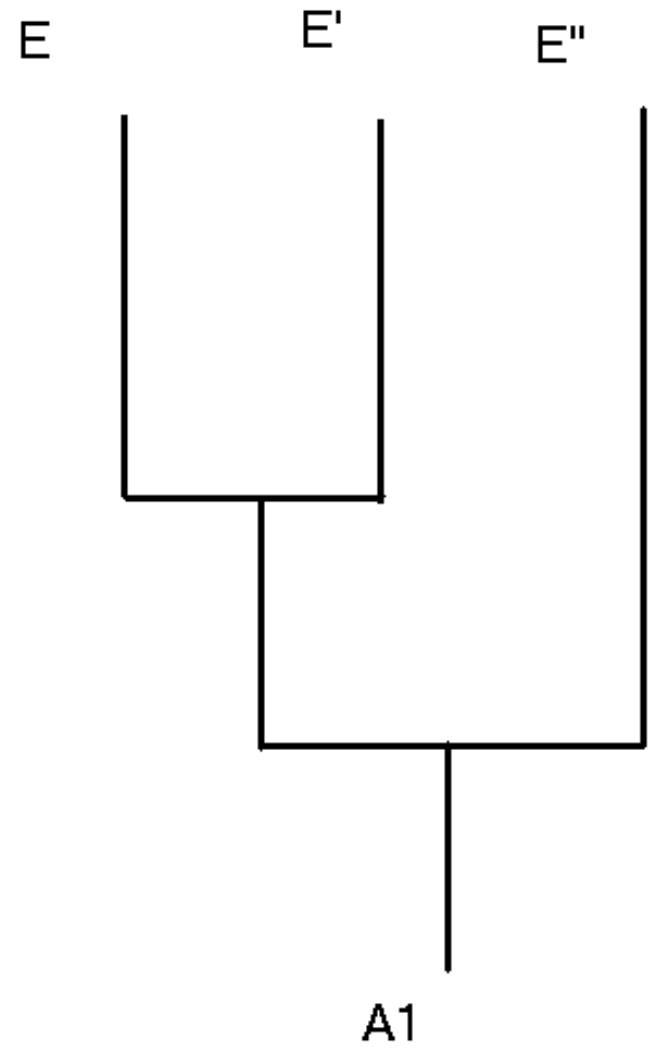

Fig. 1

Surely it is, despite the fact that there will be many more sequences in which events of kind *E* figure, with relative frequency 1/3. The reason is that $A_1$ will not count each such sequence as equiprobable, just because he will count the probability of each successor, given a single symmetric division, as equal to 1/2. In fact, given an arbitrary system of bifurcation, involving a large number of steps *N*, if *N*-step histories are equiprobable, then A at time *t* can hardly know from this how to weight his successors at time $t_k < t_N$, for that will depend on all subsequent divisions at times $t_{k+1}, \ldots, t_N$.

Working up from the local level, we arrive at the Bernouilli theorem: the probabilities of histories with the wrong statistics becomes vanishingly small, as the number of trials becomes large. So $A_1$ should surely anticipate that his likely successor will record the frequency of *E* as 1/4. This is not quite the same as saying that A believes that the frequency of *E* will be approximately *p*; but replacing "believes that" with "anticipates seeing", L2 is clearly satisfied. With the minor modification to L3, that we replace "the frequency of *E*" by "a frequency of *E* recorded by one or another of $A_1$'s survivors", we can satisfy this as well.

Is there a more charitable reading of L3, sympathetic to Loewer's purposes? We can certainly modify L3 to rule out the relational theory. Require instead that by "the frequency of *E*", Loewer does not mean a recorded frequency, but the total number of events of type *E*, no matter the histories in which they occur, divided by the total number of events. Understood in this way, the criterion, denote L3*, does rule out the models we have studied.[12]

It is not much in the way of charity. As Lewis and others have taught us, chance had better relate to credence. We give credence to what concerns us, our possible fates. Probabilities here, as in quantum mechanics, had better relate to what we care for, and to what we can perceive. So it is the probabilities for histories that matter, and the statistics that they encode, for that is all we can observe. Loewer fully agrees: his own preferred solution is entirely concerned with frequencies, as recorded by what he calls “continuing minds”, individuals for whom there is a fact of the matter as to what will be.

# 5. Reduplication

Loewer’s CMV or “continuing minds view” is another example of the strategy mentioned in connection with Deutsch and Lockwood. But it is differently motivated, as we have seen, and the arguments for it apply more generally, outside of the interpretation of probability. Let us see how.

Some terminology will be useful. Given the 4-dimensional framework, the term “individual” will be used as a neutral term, standing variously for things, worlds, minds, or events, to be modeled as space-time objects in one of two ways. According to *minimalism*, they are generally localized in time (as well as in space). Whatever larger-scale structure can be imputed to them follows as a consequence of the dynamics (realized, in quantum mechanics, as a system of correlations, of greater or lesser extent). According to *fatalism*, whilst localized, they are assured of a unique temporal history; this is to be built in to any dynamical theory that we may hold, to tell us what these histories are. The CMV, advocated by Albert and Loewer, is a version of fatalism: each mind has a unique past and future, from cradle to grave, whether or not we know what it is.

If this unique future is selected moment by moment, as with a stochastic dynamics, L1 will clearly apply; what this future is must remain unknown. But equally the probability may be epistemic, and the dynamics deterministic. Indeed L1 adds to fatalism the requirement of ignorance, and to ignorance the requirement that there be something one is ignorant of, the fate in question.

So far so good. But it now follows that there are at least as many individuals as there are possible histories. For example, given a single case of personal division, there must already have been two individuals present *prior* to the division. If not, there there could be no 1:1 map from individuals to global histories. In a sequence of $N$ divisions, there must be $2^N$ individuals present initially, all qualitatively exactly the same up to that time.

Call this mechanism *reduplication*. Many who champion Everett’s approach favor fatalism, and are prepared to pay the price of reduplication: Deutsch (1985), and Albert and Loewer (1988) are prime examples. The approach also has its adherents in classical metaphysics: witness Lewis’s response to Parfit’s arguments (Lewis 1983a). But if the relational theory is ontologically extravagant, that is nothing in comparison to reduplication, in quantum mechanics. There it amounts to replacing the universal state by an impure mixture, the convex sum over all possible histories. Unless a unique basis is to be picked out in this way, then all possible mixtures, with respect to all possible bases, had better be included as well, each as a separate plurality of histories. These are very large infinite cardinals. We shall see this explicitly in Section 8.

If that is not bad enough, a form of reduplication arises even in classical deterministic physics *without* any division of persons. Consider again the notion of "passage", and the related idea that the whole of me is at each time. If it is true that I at $t_1$ will become I at $t_2$ , then there will be no I at $t_1$ - unless, that is, an I at $t_0$ made passage through time to arrive at $t_1$. As a result, my life must be lived through an arbitrarily large number of times. In terms of Weyl's metaphor, if my awareness has left $t_1$, and crawled up the life-line of my body to $t_2$, there need be no mere automaton at $t_2$ if another awareness which was at $t_0$ has migrated to $t_1$ in the meantime.This is reduplication in the purely temporal context. Relationally, of course, there is also a duplication; there is an arbitrarily large number of person-stages. But in this plurality, and the relations among them, is already to be found the life as it is lived; there is no further duplication, where each person-stage must live through all of its moments.

Reduplication is a mistake; I take it that that is self-evident. It is to insist on the application of a concept in a context where it has already been taken into account. But in the context of the models we have considered, fatalism implies reduplication; and fatalism, urge Albert and Loewer, is necessary to make sense of probability. But that argument we have shown is mistaken.

# 6. *A Priori* Probability and Dynamics

The idealizations of the models of Section 4 are not entirely harmless; in each case it may be objected that the concept of probability is parasitic on the concept of equiprobability, defined on *a priori* grounds of symmetry. This is clearly true in the case of personal division.

*Contra* Albert and Loewer, this is already to cede that there is no fundamental obstacle to applying the notion of probability to Everett's framework; rather, the claim is that we cannot obtain the probabilities that we want, that it is not enough to appeal to an algorithm (the quantum formalism) and to merely specify the numbers (the transition probabilities).

I earlier remarked on the parallel situation in the case of state-reduction theories, where numbers alone, the modulus square of the norms, make for the difference in the actual outcomes of experiments. Evidently in neither case can we "prove" that the identifications in question are "correct". But what would count as a proof of this kind? What, indeed, would count as proof in the case of time? There is nothing *self-evidently* temporal about a 4-dimensional representation of space and time, or about space-time itself. So how do we establish that the integral of the metric along a time-like curve is "really" time, or can legitimately be understood as time?

The question is clearer in the case of space. Here there may be an antecedently given understanding of 3-dimensional space, for example, space as Euclidean space. One might then urge, on being presented with a more general manifold, that it could only be interpreted as space (or be counted a model for physical space) insofar as it has some at least of the properties of 3-dimensional Euclidean space. For example, one might require that it have constant curvature, or be locally Euclidean. Similarly, with Gödel (1949) and Savitt (1994), we might say that a space-time geometry could only "really" describe time, so long as it admits no closed time-like curves (closed loops in time).

I am not concerned to defend these particular conditions; they are simply illustrations. They are reasons to say of a theory that it is of sort rather than another (for example, whether it is deterministic or stochastic), but not to believe that the concepts of the theory apply to the world. That would amount to a criterion of the *truth* of the theory. But what could count as a criterion of truth, over and above experimental test? *In practice* there is no problem; the relational theory underwrites the standard, textbook interpretations, with use of the measurement postulates (in particular the projection postulate, where applicable, is an instance of relativization). So the theory is eminently testable (I shall come on to the details of this later). But then, to accept the adequacy of a range of concepts, as fitted to an empirical subject matter, is to accept the empirical adequacy of the theory. Conditions of sufficiency, for the application of concepts, would amount to conditions of sufficiency for the application of the theory; and this would be to insist on a condition that the theory is true.

All the same, the concept of probability presents special problems. Like arithmetic and geometry, it has a basis in *a priori* truth, in this case the theory of games. Historically, this theory was grounded on an *a priori* notion of equiprobability, defined in terms of the symmetries of the rules of the game. Like geometry, but unlike arithmetic, it was subsequently developed as an "applied" science, first in the context of classical statistical mechanics, and then to describe Brownian motion. In the process it was subject to mathematical development as well, and couched in terms of the theory of measurable spaces, and the ideas of Borel and Lebesgue. But, and here a difference with geometry, the shift in the theoretical basis of the subject was overshadowed on the empirical side, with the discovery of radioactive decay, and by the '20s, with the development of quantum mechanics. It was clear that in physics we were seeing something entirely unexpected. And it was just at this time, in the first quarter of the 20th century, culminating in the theorems of Von Neumann, Kolmolgorov, and Birkhoff, that the mathematical foundations of the modern theory of probability were perfected. The difference, between the classical and modern theory, went largely unremarked. The more so, since the old idea of *a priori* equiprobability was given a new lease of life in the theory of spin, and more generally, in the study of systems with finite-dimensional state-spaces.

This is how the models we have looked at give a false impression. They suggest that experiments have as many outcomes as there are distinct eigenvalues of the operator measured; that there is an *a priori* measure available, here as in the case of the old-fashioned theory of games, given by the cardinality of the event space. That is far from the case; every different configuration of the macroscopic apparatus - and there are uncountably many - counts as a different final state of the system. True, macroscopic events can be individuated in terms of the number of divisions on a dial, or the two sides of the coin, or the shape of the dice; but only in an interest-relative way (why is it the *side* on which the coin relevant, rather than its *position*?). And it is not for these reasons (or these are only *some* of the reasons) why the probabilities are the same for each outcome (if they *are* approximately the same). In the classical case, what really matters, in real games of cards and dice, is the dynamics, the way the cards are shuffled and the dice is thrown. In the right circumstances, the macroscopic symmetries approximately line up with the predicted probabilities; the latter, the real probabilities, are calculated on foundations which owe nothing to the concept of the equiprobability of anything.

It was recognized early on that combinatorics provide an uncertain guide to physical probability. Boltzmann developed a combinatorial method for the calculation of thermodynamic probabilities; he considered the number of "complexions" or fine-grained phases of a system, consistent with a given thermodynamic phase; but he was at pains to justify this method of computation by appeal to the H-theorem, and to the concept of *Stosszahlansatz* (in effect an equilibrium assumption governing molecular collisions).

Einstein was even more critical of the use of combinatorics, and of the use of any *a priori* concept of equiprobability, insisting instead on the use of a measure on phase space linked directly to the dynamics.[13] The details are instructive. Einstein identified the probabilities of particular states or phase-space regions $A \subset \Re$ with the limiting values of the average ratio of time that the system spends in the region $A$. If $U(t, x_0) = x(t) \in \Re$ is the phase-space point at time $t$, representing the system with initial data $x_0$ at $t = 0$, then the average time in $A$ is

$$\mu(A) = \lim_{t \to \infty} \frac{1}{\tau} \int_0^t \chi_A(U(t, x_0)dt \qquad (2)$$

where $\chi_A$ is the characteristic function of $A$. If the RHS exists, independent of the initial data $x_0$ at $t = 0$, we have a measure $\mu$ on phase space independent of time. For Lebesgue-Stieltjes measurable sets, we can then define a probability density $\rho$ on $\Re$ as follows:

$$\mu(A) = \int_A d\mu = \int_{\Re} \chi_A(x)\rho(x)dx \qquad (3)$$

From a physical point of view the difficulty is to prove that the RHS of Eq.(2) does exist independent of $x_0$ . This was, in effect, the point of Boltzmann's hypothesis of molecular disorder. Einstein argued for this result on the assumption that there do not exist any constants of motion other than the energy (so, in particular, no characteristic function for any $A$ can be be constant in time and the system must explore all of phase space). The existence and uniqueness of $\rho$ was eventually proved by von Neumann in a slightly weakened form (the mean ergodic theorem); we owe the stronger version to Birkhoff (1931). Both Von

Neumann and Birkhoff made routine use of Borel's methods (the basis of the Lebesgue theory of measure). Accordingly, what was proven was uniqueness and existence, for almost all points $x_0$ , given that the only measurable subset of $\Re$ invariant under the time evolution is the whole of $\Re$ (this is Birkhoff's versions of Einstein's hypothesis). The further question, of whether the latter notion of ergodicity (or "metric indecomposability") actually holds, for systems of physical interest, has only been answered in special cases, for example hard-body impacts with no long-range forces;[14] the important point, for our purposes, is that whether or not it holds is a matter of the detailed dynamics. There is no clear concept of physical probability, in the classical theory, independent of dynamics.

Measure theory and dynamics define equiprobabilities, if any. We do not have a definition of any partition of phase space into complexions and phase-space cells, without the use of the concept of measure; this is so even when we divide phase-space into cells according to powers of Planck's constant. There is no physically meaningful notion of probability which is not underwritten by the dynamics, quantum or classical. As we know from the non-relativistic solutions to the problem of measurement, the state-reduction and de Broglie-Bohm theories, it is position and center of mass variables that are observed, which always have a continuous spectrum.[15] There is no *a priori* way to individuate experimental outcomes on the basis of the position of a needle on a dial; on the contrary, it is the application of the theory to the apparatus that tells us how to mark out the divisions in the first place. What counts as counting is up to the measure to say. That is what is wrong with Deutsch's and Lockwood's view, that probabilities must be interpreted in terms of the cardinalities of sets of identical individuals. Such cardinalities can only be defined by the measure, not the other way round.

But isn't the ergodic theorem an example of a "proof" that the concept of probability can be applied to the world? Isn't this what I have been saying cannot be provided? Indeed it is not - or if it is, then there is an equally conclusive proof that the relational theory of probability can be applied to the world. For the ergodic theorem requires the infinite time limit; over a finite time, the proportion of the time that the system has property $A$ may or may not be related to $\mu(A)$. Given ergodicity, we no more have a reduction of probability to mean length-of-stay, than we have a reduction of probability to relative frequencies by appeal to the Bernouilli theorem.[16] It is only in the infinite limit that the two can be identified.

Probability cannot be reduced to anything else: that is what I have been saying throughout. But we have learned something too about the nature of justification. It is a mistake to conclude from this history that the question of ergodicity is a red-herring, that we learn nothing to justify the use of the measure æ as probability; equally, it would be perverse to deny that the Bernouilli theorems provide a link with statistics. Rather, these arguments are exemplary of how the concept of probability is linked to evidence, of what justification for it consists in.

# 7. Probability and Observation

I have said that the relational theory of probability can make sense of experimental practice and that the evidential basis of the theory is left intact. This is fundamental to the justification of the theory, and without it we have nothing at all. But there is an intuitive difficulty on just this score, related to the usual problem of statistical theories, the problem of what is to count as falsifying evidence. In the relational approach this difficulty is particularly acute.

The difficulty is this. There is no obvious sense to the notion of a prediction - that the optical interference fringe will be such-and-such - since by the lights of the theory this statement must be relativized. Such-and-such relative to what? Relative to $A$ the fringe system $F$ is as it should be; relative to $A'$ it is $F'$, say a fringe system which is statistically anomalous. So the prediction is both confirmed and disconfirmed.

Evidently, we need to take account of tense (the fact that $F$ is *predicted*, it lies in the *future*). But it is equally clear that, viewed in that way, what the theory says is that the probabilistic relations are such-and-such, and we have yet to get clear about how such statements are tested. We see a similar difficulty arises in every statistical theory, including stochastical dynamical theories. For a finite number of trials, i.e. for finite photon number densities, the right interference fringe is only highly probable, not certain. But if we try to make *that* the prediction, i.e. we say “the right sort of interference fringe is highly probable”, then we are back to the problem of how that in turn is to be tested.

If we view probability as fundamentally relational, and at the same time as irreducible to other physical quantities, then at some level or other we had better be able to recognize probability in the evidence itself. Our guide as ever is the space-time theory. What is a prediction, viewed in 4-dimensional space-time terms? It is presumably a statement, made at one event, which purports to describe a later event. How is this to be verified? How are we to determine the relations between the two events? We want to say: we wait and see, and we will learn the result in due course. The question is then what to make of these claims, and of how they are in their turn to be tested.

Whatever one’s views on the geometric description of time, in one context at least it is quite clear that we can relate it to observation, namely to the perception of time-relations in the case of *local coincidences*. As Einstein put it, considering the time of arrival of a train, and the position of the hand of his watch, these coincidences “do not come into question” in his theory. Any ambiguities concerning them are to be “surmounted by an abstraction” (Einstein 1905 p.892); were it otherwise it would be hard to imagine how the theory could be tested.

My proposal is this: probabilistic relations can be read off from local perceptions. They concern the here-and-now, and the probable. In particular, we can see relations between contemporary records, of the past, and what is presently on show. Judging the records reliable, we infer to the past; if the present is too unlikely, according to the quantum mechanical probability relative to that past, then quantum mechanics is to be rejected. Any ambiguity here is to be "surmounted by an abstraction".

Is this conclusion surprising? If any probabilistic theory can be tested (state reduction theories, for example), there must come a time, for each one of us, when we decide whether that theory has or has not passed an observational test.[17] For the test to be decisive, it had better have involved a large number of trials. Whether or not the number is large enough (what counts as simultaneous) is a matter to be "surmounted", i.e. we are to vote with our feet. These claims are surely mundane.

In effect the proposal is that we assume that records of the past are reliable (for example memories, under usual conditions), and that evidence for present states of affairs are reliable (perception, in normal circumstances). The one had better square with the other, as specified by the theory under test; if not the theory is rejected. Obviously, I may choose instead to doubt occurrent perceptions (I can't believe my eyes), or to doubt the records of the past (I don't trust my memory), or to doubt that what I recall was testimony to anything (I must have been hallucinating); but again, all of this is familiar ground.

If this is our account of experimental test, it remains to be seen whether, on the relational approach, quantum mechanics predicts that it will be confirmed. But so it does: relative to a record of some past prediction, correctly made, the present will probably confirm it. But does it not also predict its falsification? Since all possibilities are in some sense realized, including records of anomalous statistics, that would seem to follow. But anomalous states of affairs, albeit that they exist, have vanishingly small measure (relative to any other state of affairs whatsoever); there likewise exist anomalous worlds in the theory of Deutsch, and anomalous "continuing minds" in the theory of Albert and Loewer, reduplication and all.

In point of fact, the relational view is in this respect better off than either alternative. For it is committed to no more than the existence of the anomalous records, not to the histories that they purport to record. For, on the relational view, a historical event $H$ has only happened, relative to a state of affairs $A$, insofar as relativizing to $A$ one really does pick out $H$ (with the amplitudes for events of a different type, say $H'$, close to zero). If $A$ contains anomalous records, the probability of very different past events (involving fraud or whatever) becomes much higher. In the limit, in which $A$ has no coherent structure at all, from a quasi-classical point of view, then neither has it a determinate past (for there will be nothing in the past with which it is, as a matter of the unitary dynamics, strongly correlated). Its probable future, meanwhile, is invariably that in which the statistics are normal. (One would like to say that the *number* of deviant histories is vanishingly small, in comparison to normal histories, and forego use of the quantum mechanical measure altogether; but that is to return to our old *a priorist* ways.[18])

Is probability then something subjective, a matter of records and the like? Not at all: these relations exist as much in the here and now, over the specious present, where memory and perception fuse. As such they are familiar to us, in terms of the momentary sense of passage. This is probabilistic change; with probability close to 1, my experience is at each moment predicted to be "normal", in the statistical sense; and I test this claim within the moment. And that would remain the case, even given a *stochastic* dynamics.

# 8. Reduplication and Superposition

Decoherent histories theory has hitherto played only a tacit role.[19] To see more precisely what is involved in adopting fatalism, and with that reduplication, we need to make use of it explicitly. The conclusion, as already announced, is that the additional histories, introduced on the assumption of reduplication, are superposed, according to minimalism. Correspondingly, the fatalist made use of a mixed state, obtained by taking the convex sum over reduplicated histories, whilst for the minimalist the state remains pure.

First some notation. Histories are represented by ordered products of projections, of the form:

$$C_{\underline{\alpha}} = P_{\alpha_f}\ldots\ldots P_{\alpha_1}P_{\alpha_0}P_{\alpha_{-1}}\ldots P_{\alpha_{-p}} = C_{\underline{\alpha}_f}P_{\alpha_0}C_{\underline{\alpha}_p}.$$

Here $\underline{\alpha}_{f=\alpha_f}\ldots\alpha_1$ and $\underline{\alpha}_p = \alpha_{-1}\ldots\alpha_{-p}$ are, respectively, a particular future (from $t_1$ to $t_f$ ), and a particular past (from $t_{-1}$ to $t_{-p}$ ), relative to $t_0$. The condition of "medium decoherence", in Gell-Mann and Hartle's sense, requires that interference effects between distinct histories are vanishingly small. This condition amounts to this:

$$Tr(C_{\underline{\alpha}}\rho C^{\dagger}_{\underline{\alpha}'}) \approx Tr(C_{\underline{\alpha}}\rho C^{\dagger}_{\underline{\alpha}})\delta_{\underline{\alpha},\underline{\alpha}'}. \qquad (4)$$

When $\underline{\alpha}=\underline{\alpha}'$ we have the absolute measure of the history $\underline{\alpha}$ in the universal state $\rho = |\ \Psi >< \Psi\ |$. But the quantities ordinarily of interest to us are conditional probabilities. These are defined in the obvious way; for example, the probability of the present $\alpha_0$, conditional on the past $\underline{\alpha}_p$, is:

$$\mathrm{Prob}(\alpha_0/\underline{\alpha}_p) = Tr(P_{\alpha_0}C_{\underline{\alpha}_p}\rho C^{\dagger}_{\underline{\alpha}_p}P_{\alpha_0})/Tr(C_{\underline{\alpha}_p}\rho C^{\dagger}_{\underline{\alpha}_p}).$$

Given fatalism, we can suppose that probabilities all derive from the measure on the space of complete histories. These histories can be thought of as "worlds", in Lewis' sense; indeed, we can take over his theory of probability more or less unchanged.
For the probability of events $\alpha_k$ :

> We may picture the situation as follows. The partition divides logical space into countless tiny squares. In each square there is a black region where à holds and a white region where it does not. Now blur the focus, so that divisions within the squares disappear from view. Each square becomes a grey patch in a broad expanse covered with varying shades of grey. Any maximal region of uniform shade is a proposition specifying the chance of à . The darker the shade, the higher is the uniform chance of à at the world in the region. The worlds

> themselves are not grey - they are black or white, worlds where à holds or where it doesn't - but we cannot focus on single worlds, so they all seem to be the shade of grey that covers their region. (Lewis 1983b p.99-100)[20].

Against this, on the minimalist view, there is no fact of the matter as to what the past is, over and above what is correlated with the present. Nor is there any fact of the matter as to the future, subject to the same proviso. For all that, given decoherence (and *only* given decoherence), the difference between fatalism and minimalism makes no difference to the probabilities.

We can see this as follows. Given minimalism, the probability of $\underline{\alpha}_f$ relative to $\alpha_0$ is

$$\mathrm{Prob}(\underline{\alpha}_f/\alpha_0) = Tr(C_{\underline{\alpha}_f} P_{\alpha_0} \rho P_{\alpha_0} C^{\dagger}_{\underline{\alpha}_f}) / Tr(P_{\alpha_0} \rho P_{\alpha_0}). \tag{5}$$

The fatalist, coming complete with a definite past, will conditionalize on that as well as on the present. But not knowing which his past is, he must take into account all of them, weighted according to which is most likely his, conditional on the present. He will therefore compute:

$$\sum_{\underline{\alpha}_p} \mathrm{Prob}(\underline{\alpha}_f/\alpha_0 \underline{\alpha}_p)\mathrm{Prob}(\underline{\alpha}_p/\alpha_0)$$

$$= \sum_{\underline{\alpha}_p} \left( \frac{Tr(C_{\underline{\alpha}_f} P_{\alpha_0} C_{\underline{\alpha}_p} \rho P^{\dagger}_{\underline{\alpha}_p} P_{\alpha_0} C^{\dagger}_{\underline{\alpha}_f})}{Tr(P_{\alpha_0} C_{\underline{\alpha}_p} \rho C^{\dagger}_{\underline{\alpha}_p} P_{\alpha_0})} \cdot \frac{Tr(P_{\alpha_0} C_{\underline{\alpha}_p} \rho C^{\dagger}_{\underline{\alpha}_p} P_{\alpha_0})}{Tr(P_{\alpha_0} \rho P_{\alpha_0})} \right)$$

$$= \sum_{\underline{\alpha}_p} \mathrm{Prob}(\underline{\alpha}_f \alpha_0 \underline{\alpha}_p/\alpha_0) = \mathrm{Prob}(\underline{\alpha}_f/\alpha_0) \tag{6}$$

i.e., he obtains the same result as the Minimalist, Eq.(5). Here decoherence enters in the definition of the LHS, and in the last step on the RHS.

Analogous calculations for the probability of a single future event, or for past events, are trivial; contrary to what might be expected, the difference in the two views leads to no difference at the level of the probabilities that really matter. The Fatalist can hardly object that "really" the probability for $\underline{\alpha}_f$ is one of the summands in Eq.(6) (because "really" his history is one of the $\underline{\alpha}_p$ 's); in truth the probability is either 0 or 1, for his *real* history, by his lights, includes a complete future as well.

What if we did not have decoherence? Decoherence, recall, is required for the definition of retrospective probabilities. But the fatalist can still define a notion of probability for the past, conditional on the present (denote 'chance'). For according to Fatalism, the chance of $\underline{\alpha}_p$ given $\alpha_0$ is surely the absolute chance of $\alpha_0 \underline{\alpha}_p$, divided by the sum of chances for all the possible ways $\alpha_0$ might have come about. The latter is just the sum of chances for all histories terminating in $\alpha_0$ , i.e. the quantity:

$$\mathrm{Chance}(\alpha_0) = \sum_{\underline{\alpha}_p} Tr(P_{\alpha_0} C_{\underline{\alpha}_p} \rho C^{\dagger}_{\underline{\alpha}_p} P_{\alpha_0}).$$

We then obtain, for the retrodictive chance:

$$\text{Chance}(\underline{\alpha}_p/\alpha_0) = \frac{Tr(P_{\alpha_0}C_{\underline{\alpha}_p}\rho C^{\dagger}_{\underline{\alpha}_0}P_{\alpha_0})}{\sum_{\underline{\alpha}'_p} Tr(P_{\alpha_0}C_{\underline{\alpha}'_p}\rho C^{\dagger}_{\underline{\alpha}'_p}P_{\alpha_0})}$$

This expression, unlike quantities of the form:

$$\frac{Tr(P_{\alpha_0}C_{\underline{\alpha}_p}\rho C^{\dagger}_{\underline{\alpha}_p}P_{\alpha_0})}{Tr(P_{\alpha_0}\rho P_{\alpha_0})}$$

sums to unity (over $\underline{\alpha}_p$) whether or not the decoherence condition is satisfied, so is formally admissible as a notion of probability. Proceeding as before, the fatalist obtains

$$\sum_{\underline{\alpha}_p}\left(\frac{Tr(C_{\underline{\alpha}_f}P_{\alpha_0}C_{\underline{\alpha}_p}\rho P^{\dagger}_{\underline{\alpha}_p}P_{\alpha_0}C^{\dagger}_{\underline{\alpha}_f})}{Tr(P_{\alpha_0}C_{\underline{\alpha}_p}\rho C^{\dagger}_{\underline{\alpha}_p}P_{\alpha_0})}\cdot\frac{Tr(P_{\alpha_0}C_{\underline{\alpha}_p}\rho C^{\dagger}_{\underline{\alpha}_p}P_{\alpha_0})}{\sum_{\underline{\alpha}'_p} Tr(P_{\alpha_0}C_{\underline{\alpha}'_p}\rho C^{\dagger}_{\underline{\alpha}'_p}P_{\alpha_0})}\right) = \frac{\sum_{\underline{\alpha}_p} Tr(C_{\underline{\alpha}_f}P_{\alpha_0}C_{\underline{\alpha}_p}\rho P^{\dagger}_{\underline{\alpha}_p}P_{\alpha_0}C^{\dagger}_{\underline{\alpha}_f})}{\sum_{\underline{\alpha}'_p} Tr(P_{\alpha_0}C_{\underline{\alpha}'_p}\rho C^{\dagger}_{\underline{\alpha}'_p}P_{\alpha_0})}$$

Evidently the RHS is exactly the expectation value for $\underline{\alpha}_f$, conditional on $\alpha_0$, using the impure mixture:

$$\rho_{mix} = \sum_{\underline{\alpha}_p} C_{\underline{\alpha}_p}\rho C^{\dagger}_{\underline{\alpha}_p}.$$

Only given (exact) decoherence, i.e. Eq.(4) (with strict equality), do we obtain the same quantities as for the pure state $\rho$. The Fatalist will not obtain Eq.(5), so will disagree with the minimalist as to chances for future events.

What concerns us, as creatures made of biochemical processes, are decoherent histories. In that case, if $\rho_{mix} \approx \rho$, why not opt for the uncontentious notion of probability, and embrace fatalism? It is true that the price is reduplication, but the step would seem to be natural and innocuous enough; if it makes for an integrated totality in purely physical terms - in the way that the totality of hyper-surfaces of some space-time foliation makes for a natural and integrated totality, namely space-time - the price may well be worth paying.

But we do not attain an integrated totality in this way. On the contrary, what makes the universal state a unified object is exactly all of the phase relationships between its various histories; just what is annulled by the Fatalist. This is obvious given medium (approximate) decoherence, as is physically more realistic; in that case the probabilities are only approximately the same. Nor is the step innocuous; it makes all the difference in the world to the preferred basis problem. If we count as a plurality all the histories of one decoherent history space, represented by $\rho_{mix}$, then what of other decoherent history spaces? And histories which do not decohere? What now of the choice of measure?

The relational theory is by contrast austere. The familiar intuitions of probability are not needed, for we have learned how to replace them with ones better suited to relativity. The universal state, with its unitary evolution, is a single object in its own right. We may take various cross-sections through this object, and consider relations among various of its parts, but the totality of cross-sections and relations does not exist as something over and above the original. I take it that this is the guiding inspiration of the relational approach, and the core concept of the physics.

# Postscript (August 2001)

There is an ambiguity in Section 5 that, at the time of writing, I thought was harmless: whether by "individual" I meant something localised in time (although this is what I meant). I was then led to make a downright error: I should not have cited Lewis's 1983a as an example of fatalism. Although Lewis is a fatalist at the level of worlds - so my comments in Section 8 stand unchanged - he is not committed to this thesis at the level of persons. His (1983a) response to Parfit was, however, concerned only with persons, and there he explicitly allowed that person stages may be shared. In that case a person stage (an individual in my sense) need not have a unique future or past. In fact, in this essay, Lewis offers an account of identity that may well be suited to the miminalist.

# Footnotes

* I would like to thank Hilary Putnam, David Albert, and Paul Tappenden, for stimulus and helpful criticism.
1. For example, histories can only incorporate reliable records if they decohere (Halliwell 1994); complex adaptive systems, which satisfy a criterion of "fitness", must decohere (Saunders 1993).
2. It is also more difficult, given the genetic approach to the preferred-basis problem just sketched. Wiith Dowker and Kent (1995), we may suppose that only a *single* history is selected from a *particular* history space, or from each history space, so there is no problem of probability: but then the nature of "the observer" is irrelevant to the preferred basis problem (*which* history space?), and the latter must be solved on the basis of new physical principles.
3. Saunders (1995, 1996a); near neighbors, apart from Everett, include the unitary theories of Zeh, Gell-Mann and Hartle, Zurek, Vaidman, and Rovelli. There are also similarities with Merman's 'Ithica' interpretation (Merman 1998), his rhetoric against the 'many worlds extravaganza' notwithstanding.
4. It may be objected that the general theory of relativity makes no such claim. For a discussion of this and the other points just raised I refer to Saunders (1996a).
5. The same problem arises in the case of tense. McTaggart (19xx) argued that so-called "A-determinations", words such as "present", "past", "future", are inconsistent just because every event is past, present, and future; but from a relational point of view, two events $x$, $y$ can both be present, and one event be both past and future, so long as "present" etc. are in relation to distinct events $z$.
6. This stipulation eliminates by fiat the possibility of recombination of branches; in this and other respects the model is idealized. (See Section 6.)
7. Specifically, if $P$ is a projection and $\rho$ is a density matrix on a complex separable Hilbert space $H$, dim($H$)$\leq$ 3, the expectation of $P$ is of the form $Tr(P\rho) = \sum_k < \varphi_k, \rho\varphi_k >$, where $\{\varphi_k\}$ is any orthonormal basis on $H$ and '$Tr$' is the trace. In the pure case, $\rho = | \Psi >< \Psi |$; this yields $< \Psi, P\Psi >$, as required by the Born rule.
8. Weak and strong versions have subsequently been proved, and with greater rigor. See e.g. Fahri *et al* (1989).
9. This mechanism should be distinguished from reduplication (Section 5), for it conerns only the question of whether or not probability and measure must be interpreted in terms of the cardinality of some underlying set of equiprobable individuals. Against this, see Section 6.
10. There remain *microscopic* indeterminate events, of course. Also, if we relativized to the here as well as the now, there will exist indeterminate space-like events even on the macroscopic level. However, decoherence time-scales are so short that correlations with superposed states turn into superpositions of correlated states more or less instantaneously. (For some this is not enough; see Tappenden 1998.)
11. Symmetries of the human brain do, of course, allow a certain realization of Parfit's scenario, namely when the corpus collusum, connecting left and right lobes of the cerebral cortex, is severed. (Other brain functions, for example those of the brain stem, remain in common, however.)
12. The *measure* of events of type $E$, as defined by quantum mechanics, is the same whether or not they are first grouped into histories. In effect the starred versions of L2 and L3 require the use of cardinal numbers of sets of events in place of the quantum mechanical measure. See fn.18.
13. Einstein (1903). It is clear that Einstein had in mind Planck's use of combinatorics in his derivation of the spectral distribution law for radiative equilibrium (the black-body law). It is a nice moment to revisit, marking as it did the very beginning of quantum mechanics.
14. For further historical background, see e.g. Pais (1982), von Plato (1994). See Sklar (1993) for a comprehensive study of more recent developments, including non-equilibrium theory and cosmology.
15. I set to one side speculation on the significance of the Hawking-Beckenstein bound, in quantum gravity. Developments on that score may yet entirely transform our problem situation, including the problem of measurement.
16. This point was missed by Einstein, as by many others; see, e.g. Pais (1982) p.67-8.
17 Obviously the great majority of us go on hearsay, but hearsay too is evidence, and evidence of the type just specified.
18. Given infinite reduplication, one does have this form of words available, defining ratios in the number of (qualitatively identical) individuals in each branch, by ratios of the measures of those branches; with that one can satisfy L3*, for example. But obviously these ratios, involving infinite cardinals, do not strictly speaking exist; they are convenient fictions.
19. Decoherence is a condition for the existence of retrodictive probabilities and, less straightforwardly, for the Bernouilli theorem. It also ensures that the relation of value-definiteness is transitive (Saunders 1996b). (See also fn.1.)
20. Lewis used the symbol '*A*', not '$\alpha_k$ ', and defines a "proposition" in terms of its extension, i.e. the worlds at which it is true. Lewis is disinclined to link his modal realism with the theory of DeWitt and Deutsch (including reduplication); but he can hardly deny that his theory of probability fits it very well. It is no surprise that fatalism eliminates what is distinctive to quantum mechanics, the superposition principle, and postulates instead a set-theoretic collective, in place of the universal state. For a detailed study of the many-worlds interpretation made out in Lewis's terms, I refer to Hemmo (1996).

Simon Saunders
Sub-Faculty of Philosophy
University of Oxford,
10 Merton St., Oxford OX1 4JJ